\documentclass[journal,onecolumn]{IEEEtran}
\ifCLASSINFOpdf
\else
\fi
\usepackage[margin=1in]{geometry}

\usepackage{amssymb,amsmath,amsthm,amsfonts}
\usepackage{mathtools}
\usepackage{enumitem}
\usepackage[numbers,comma,sort&compress]{natbib}
\usepackage{authblk}
\usepackage{graphicx}
\usepackage[font=small]{caption}

\usepackage[labelformat=simple]{subcaption}
\usepackage{float}
\usepackage[ruled,vlined,linesnumbered]{algorithm2e}
\usepackage{physics}
\usepackage{footnote}
\usepackage{xcolor}
\usepackage{bbding}

\usepackage{tikz}
\usetikzlibrary{calc,patterns,angles,quotes}
\usepackage{siunitx}
\usetikzlibrary{quantikz}

\usepackage{multirow}
\usepackage{tabularx}
\usepackage{booktabs}
\usepackage{threeparttable}
\usepackage{adjustbox}

\usepackage{hyperref}

\usepackage{extarrows} 

\usepackage{makecell}

\newtheorem{theorem}{Theorem}
\newtheorem{definition}{Definition}
\newtheorem{lemma}{Lemma}

\newcommand{\thm}[1]{\hyperref[thm:#1]{Theorem~\ref*{thm:#1}}}
\newcommand{\cor}[1]{\hyperref[cor:#1]{Corollary~\ref*{cor:#1}}}
\newcommand{\defn}[1]{\hyperref[defn:#1]{Definition~\ref*{defn:#1}}}
\newcommand{\lem}[1]{\hyperref[lem:#1]{Lemma~\ref*{lem:#1}}}
\newcommand{\prop}[1]{\hyperref[prop:#1]{Proposition~\ref*{prop:#1}}}
\newcommand{\fig}[1]{\hyperref[fig:#1]{Figure~\ref*{fig:#1}}}
\newcommand{\tab}[1]{\hyperref[tab:#1]{Table~\ref*{tab:#1}}}
\newcommand{\algo}[1]{\hyperref[algo:#1]{Algorithm~\ref*{algo:#1}}}
\renewcommand{\sec}[1]{\hyperref[sec:#1]{Section~\ref*{sec:#1}}}
\newcommand{\append}[1]{\hyperref[append:#1]{Appendix~\ref*{append:#1}}}
\newcommand{\fac}[1]{\hyperref[fac:#1]{Fact~\ref*{fac:#1}}}
\newcommand{\lin}[1]{\hyperref[lin:#1]{Line~\ref*{lin:#1}}}

\def\>{\rangle}
\def\<{\langle}

\renewcommand{\ket}[1]{|#1\rangle}

\newcommand{\E}{\mathbb{E}}

\renewcommand{\H}{\mathcal{H}}

\def\:{\hbox{\bf:}}

\makeatletter
\newcommand{\doubletilde}[1]{{%
  \mathpalette\double@tilde{#1}%
}}
\newcommand{\double@tilde}[2]{%
  \sbox\z@{$\m@th#1\tilde{#2}$}%
  \ht\z@=.9\ht\z@
  \tilde{\box\z@}%
}
\makeatother

\let\oldnl\nl
\newcommand{\nonl}{\renewcommand{\nl}{\let\nl\oldnl}}

\SetKwComment{Comment}{/* }{ */}

\SetKwInOut{Input}{Input}
\SetKwInOut{Registers}{Registers}
\SetKwInOut{Parameters}{Parameters}
\SetKwInOut{Notations}{Notations}
\SetKwInOut{Output}{Output}
\SetKw{Apply}{Apply}
\SetKw{Construct}{Construct}
\SetKw{Estimate}{Estimate}
\SetKw{Return}{Return}

\begin{document}
%
\title{An efficient quantum algorithm for \\ independence testing}

%
%
%

\author{Zongxing Xiong
}

\maketitle

\begin{abstract}
We propose an efficient quantum algorithm for testing independence of discrete distributions on \([n]\times[m]\), for any $2\leq m\leq n$. Given the purified quantum query-access of an unknown distribution $p = (p_{ij})_{i\in [n], j\in[m]}$, our quantum tester can decide whether $p = p_I \times p_J$ or $\|p - p_I \times p_J\|_1 \geq \epsilon$ with success probability at least 2/3 using
$$\widetilde{O}\!\left(\max\!\left\{\frac{\sqrt n}{\epsilon},
       \min\!\left(\frac{(nm)^{1/3}}{\epsilon^{4/3}},
                         \frac{\sqrt{nm}}{\epsilon}\right)\right\}\right)
$$
queries, where $p_I$, $p_J$ are marginal distributions of the two labels. In technical aspect, the tester compares the signs of a set of preliminary samples and the signs of new test samples after some elaborately designed marginal checks. Gapped group-testing is employed to count the sign agreements and disagreements, and an excessive discrepancy of them rejects independence of the distritution. 
\end{abstract}


%
\IEEEpeerreviewmaketitle

\medskip

\section{Introduction}
Consider a pair of labels \(I\in[n]\) and \(J\in[m]\), where \([k]=\{1,\ldots,k\}\) for
each positive integer \(k\). One can generate an outcome pair by retaining both labels
from one sample, or by taking \(I\) from one sample and \(J\) from another
independent sample. Provided the original pair has distribution
\(p=(p_{ij})_{i\in[n],\,j\in[m]}\), the second procedure indeed samples from
\(p_I\times p_J\), where \(p_I(i)=\sum_{j=1}^{m}p_{ij}\) and
\(p_J(j)=\sum_{i=1}^{n}p_{ij}\). The two procedures give the same distribution
precisely when \(I\) and \(J\) are independent.

In~\cite{batu-et-al-independence-2001}, Batu et al. conducted a study about
testing independence of distributions on rectangular domains under the classical sampling model.
In independence testing, the goal is to distinguish independence from the case in which
\[
  \|p-p_I\times p_J\|_1
    =\sum_{i=1}^{n}\sum_{j=1}^{m}\bigl|p_{ij}-p_I(i)p_J(j)\bigr|
    \geq \epsilon.
\]
The absolute values matter. The differences \(p_{ij}-p_I(i)p_J(j)\) always
sum to zero, and a discrepancy of size \(\epsilon\) can be spread across many
infrequent outcome pairs. Our algorithm detects such discrepancies by recording
which sampling procedure produced a pair and comparing that record with the
source of a new test sample at the same pair.

\subsection{Main results}
We assume purified quantum query access to \(p=(p_{ij})_{i\in[n],\,j\in[m]}\). A
unitary oracle \(U_p\) prepares a purification of \(p\) from the all-zero state, and
measuring the two outcome registers yields \((i,j)\) with probability \(p_{ij}\). Each
application of \(U_p\), \(U_p^\dagger\), or either controlled version counts as one
query.

\begin{definition}[Purified quantum query-access]\label{defn:purified-query-access}
Let \(p=(p_{ij})_{i\in[n],\,j\in[m]}\) be a joint distribution. In the model of
Gily\'en and Li~\cite{gilyen-li-2020}, the algorithm has coherent access to a unitary
\(U_p\) and its inverse \(U_p^\dagger\). Registers \(I\) and \(J\) hold the outcome
labels, and \(A\) is an auxiliary register. On the all-zero input the oracle prepares
the state
$$U_p\ket{0}_I\ket{0}_J\ket{0}_A
    =\sum_{i=1}^{n}\sum_{j=1}^{m}\sqrt{p_{ij}}\,
      \ket{i}_I\ket{j}_J\ket{\phi_{ij}}_A ,
$$
where the auxiliary states \(\ket{\phi_{ij}}_A\) for \(p_{ij}>0\) form an orthonormal
family. These states may be unknown to the algorithm. An application of \(U_p\),
\(U_p^\dagger\), or either controlled version counts as one query.
\end{definition}

Grouping the terms by row or by column gives purified access to either marginal.
Omitting terms of zero marginal mass, we obtain the expansions
\begin{equation*}\begin{split}
U_p\ket{0}_I\ket{0}_{JA} = \sum_{i=1}^{n}\sqrt{p_I(i)} ~\ket{i}_I \otimes \sum_{j=1}^{m}\sqrt{\frac{p_{ij}}{p_i}}\,\ket{j}_J\ket{\phi_{ij}}_A \triangleq \sum_{i=1}^{n}\sqrt{p_I(i)} ~\ket{i}_I \ket{\varphi_{i}}_{JA}
\end{split}\end{equation*}
and
\begin{equation*}\begin{split}
U_p\ket{0}_J\ket{0}_{IA} = \sum_{j=1}^{m}\sqrt{p_J(j)} ~\ket{j}_J \otimes \sum_{i=1}^{n}\sqrt{\frac{p_{ij}}{p_j}}\,\ket{i}_I\ket{\phi_{ij}}_A \triangleq \sum_{j=1}^{m}\sqrt{p_J(j)} ~\ket{j}_J \ket{\varphi'_{j}}_{IA},
\end{split}\end{equation*}
where $p_I(i) = \sum_{j\in[m]} p_{ij}$ and $p_J(j) = \sum_{i\in[n]} p_{ij}$ are the row and
column marginal probabilities respectively. We can now state our main result.

\begin{theorem}\label{thm:independence}
Let \(2\leq m\leq n\) and \(0<\epsilon\leq 1\). Given purified quantum query access to
\(p\) on \([n]\times[m]\), there is a quantum algorithm that accepts if
\(p=p_I\times p_J\) and rejects if \(\|p-p_I\times p_J\|_1\geq\epsilon\). In either
case the algorithm succeeds with probability at least \(2/3\) and uses
\[
  \widetilde{O}\!\left(\max\!\left\{\frac{\sqrt n}{\epsilon},
    \min\!\left(\frac{(nm)^{1/3}}{\epsilon^{4/3}},
                \frac{\sqrt{nm}}{\epsilon}\right)\right\}\right)
\]
queries to $U_p$, its inverse, or their controlled versions, as in
\defn{purified-query-access}.
\end{theorem}

Ref.~\cite{diakonikolas-kane-2016} and~\cite{gilyen-li-2020} formulate independence testing in terms of distance to the
nearest product. Write \(\delta_{\mathrm{prod}}(p)=\inf_{a,b}\|p-a\times b\|_1\), where
\(a\) and \(b\) range over distributions on \([n]\) and \([m]\) respectively, then
\[
  \delta_{\mathrm{prod}}(p)\le \|p-p_I\times p_J\|_1
  \le 3\,\delta_{\mathrm{prod}}(p).
\]
by the distance comparison of Batu et al.~\cite{batu-et-al-independence-2001}. To see the second inequality, marginalize the two distributions. Each of
\(\|p_I-a\|_1\) and \(\|p_J-b\|_1\) is at most \(\|p-a\times b\|_1\), and their sum
bounds \(\|p_I\times p_J-a\times b\|_1\). The claim follows from the triangle
inequality upon taking the infimum over \(a,b\). Thus running the earlier testers at
accuracy \(\epsilon/3\) gives their stated upper bounds for our promise at accuracy
\(\epsilon\).

\tab{independence-bounds} compares the classical sample bound and the quantum query
bounds for the promise in \thm{independence}. We use \(\widetilde{O}\) to suppress
factors polylogarithmic in \(n\), \(m\), and \(1/\epsilon\). At fixed \(\epsilon\), the
bound in \thm{independence} is \(\widetilde{O}(\sqrt n)\) for \(m\leq\sqrt n\) and
\(\widetilde{O}\big((nm)^{1/3}\big)\) for \(m\geq\sqrt n\). For fixed $m$ it is
\(\widetilde{O}(\sqrt{n}/\epsilon)\). On an $n \times n$ square domain it becomes
\(\widetilde{O}\big(\min\{n^{2/3}/\epsilon^{4/3},
                n/\epsilon\}\big)\).
The same tester attains these bounds throughout the parameter range.

\medskip
\begin{table}[H]
\centering
\small
\setlength{\tabcolsep}{4pt}
\begin{tabular*}{\linewidth}{@{\extracolsep{\fill}}llll@{}}
\toprule
Tester & Resource counted & General upper bound & \(m=n\), fixed \(\epsilon\) \\
\midrule
Diakonikolas and Kane~\cite{diakonikolas-kane-2016}
  & Independent samples
  & \(\displaystyle O\!\left(\max\!\left\{
       \frac{n^{2/3}m^{1/3}}{\epsilon^{4/3}},
       \frac{\sqrt{nm}}{\epsilon^2}\right\}\right)\)
  & \(O(n)\) \\
\addlinespace[0.6ex]
Gily\'en and Li~\cite{gilyen-li-2020}
  & Purified queries
  & \(\displaystyle \widetilde{O}\!\left(\frac{\sqrt{nm}}{\epsilon}\right)\)
  & \(\widetilde{O}(n)\) \\
\addlinespace[0.6ex]
This work (\thm{independence})
  & Purified queries
  & \(\displaystyle \widetilde{O}\!\left(\max\!\left\{
       \frac{\sqrt n}{\epsilon},
       \min\!\left(\frac{(nm)^{1/3}}{\epsilon^{4/3}},
                   \frac{\sqrt{nm}}{\epsilon}\right)\right\}\right)\)
  & \(\widetilde{O}(n^{2/3})\) \\
\bottomrule
\end{tabular*}
\caption{Upper bounds for testing independence for \(2\leq m\leq n\) and
\(0<\epsilon\leq1\).}
\label{tab:independence-bounds}
\end{table}

\subsection{Techniques}

We apply the split-distribution construction of Diakonikolas and
Kane~\cite[Definition~2.4 and Fact~2.5 of the full version]{diakonikolas-kane-2016} to
each marginal. Samples from \(p\) determine the number of copies of each row and
column. On subsequent samples we choose a copy of each label uniformly, making the two
choices independently. If row \(i\) has \(a_i\) copies and column \(j\) has \(b_j\)
copies, each of the \(a_i b_j\) refined pairs has probability \(p_{ij}/(a_i b_j)\)
under the joint distribution and \(p_I(i)p_J(j)/(a_i b_j)\) under the product of its
marginals. The absolute discrepancies over these copies sum to
\(\lvert p_{ij}-p_I(i)p_J(j)\rvert\). Splitting therefore reduces the mass of frequent
labels and preserves the distance in the promise.

On the refined domain, each preliminary sample comes from one of four equally likely
choices. A sample from the joint distribution carries sign \(+1\), and a sample from
the product of its marginals carries sign \(-1\). The remaining choices assign signs
\(+1\) and \(-1\), respectively, to a uniformly sampled outcome pair. Uniform samples
give every pair a positive probability of appearing and contribute equally to the two
signs. For each observed pair \(z\), the tester records the first sign as \(G(z)\).

The tester uses the same four choices for each new test sample. It checks the row and
column against separate lists drawn from the corresponding refined marginals. On each
axis it either selects no addresses or takes the first \(k\) addresses of a uniformly
shuffled list, choosing \(k\) independently and uniformly from the integers between
zero and the list length, inclusive. The label passes if none of the selected entries
contains it. If both checks pass and the pair has a recorded sign, the sample
contributes to \(N_+\) when the signs agree and to \(N_-\) when they disagree.

Fix the split and let \(\mu_z\) be the mean sign conditional on the refined pair \(z\).
Under independence, \(\mu_z=0\) for every \(z\). Hence \(N_+-N_-\) has mean zero even
conditional on the preliminary signs, the marginal lists, and all selected addresses.
Under dependence, the preliminary sign and the new sign at \(z\) both have mean
\(\mu_z\). Conditional on observing \(z\) in the preliminary samples and again in an
independent new test sample, the expected product of the two signs is therefore
\(\mu_z^2\). Conditioning further on the recorded sign gives \(G(z)\mu_z\), which may
have either sign.

For \(D=nm\), set
\[
  d=\min\!\left\{\frac{1}{\epsilon^2},\frac{\sqrt D}{\epsilon}\right\},
  \qquad
  h=\min\!\left\{D,\left\lceil(Dd^2)^{1/3}\right\rceil\right\}.
\]
Let \(Y\) denote the expected score of one new test sample conditional on the split,
the table \(G\) of recorded signs, and both marginal lists. The score is \(+1\) for a
contribution to \(N_+\), \(-1\) for a contribution to \(N_-\), and zero otherwise. The
expectation is over the new test sample and its choice of addresses. Under
independence, \(Y=0\). If \(\|p-p_I\times p_J\|_1\ge\epsilon\),
Subsection~\ref{subsec:analysis-correctness} shows that with constant probability over
the split, preliminary samples, and marginal lists,
\[
 |Y|\ge c
 \begin{cases}
 h\epsilon^2/D,       & \epsilon\ge D^{-1/2},\\
 \epsilon/\sqrt D,    & \epsilon<D^{-1/2},
 \end{cases}
\]
for an absolute constant \(c>0\). In the first regime we use the mean sign product
\(\mu_z^2\) at observed pairs. In the second regime, \(h=D\) and both marginal checks
are skipped. The uniform component of the preliminary samples ensures that, with
constant probability, the recorded pairs account for at least half the squared
\(\ell_2\) distance between the refined joint distribution and the product of its
marginals. Conditional on the split, preliminary outcome pairs, and marginal lists, the
second and fourth moments of \(Y\) over the recorded signs then give the stated bound
on \(|Y|\).

Both counts use the same new test samples, marginal lists, and address choices. For
\(s\in\{+1,-1\}\), the quantum procedure compares \(N_s\) with successive thresholds by
querying the sample and marginal lists directly. Consider a sample \(t\) with sign
product \(s\) that fails a marginal check. Let \(H_t^*\) count its matching reference
addresses, assigning weight one to each row match and weight \(\lceil n/m\rceil\) to
each column match. The sample contributes \(1/H_t^*\) to the sum \(W_s\). We use the
same cap \(W_0\) at all thresholds. At threshold \(v\) with gap \(g\), the base
comparison returns \(\mathsf{low}\) with probability at least \(2/3\) whenever
\(N_s\le v\) and \(W_s\le W_0\). If \(N_s\ge v+g\), it returns \(\mathsf{high}\) with
probability at least \(2/3\) for every value of \(W_s\). The tester repeats the
comparisons to reduce the error probability. It rejects if either count is too large or
if the absolute difference between the estimates exceeds the threshold in
Subsection~\ref{subsec:tester-decision}.

In each round, the tester records signs from \(O(h)\) preliminary samples. On an axis
of original size \(L\), it skips the marginal check with probability \(\min\{1,h/L\}\).
Flattening the marginals and recording the signs require \(\widetilde{O}(h)\) purified
queries. To prove Lemma~\ref{lem:one-sided-count}, we use the gapped group-testing
weights of Ambainis et al.~\cite{ambainis-et-al-ggt}. All groups of sample indices
share an initial sequence of reference queries, so its cost enters the bound only once.
Subsection~\ref{subsec:analysis-queries} bounds the remaining cost by
\(\widetilde{O}(d\sqrt{D/h}+\sqrt{nd})\) queries per round, including reads of the new
test samples and both marginal lists. Since the number of rounds is constant, the total
query complexity is
\[
  \widetilde{O}\!\left(h+d\sqrt{D/h}+\sqrt{nd}\right).
\]
For \(\epsilon\geq D^{-1/2}\), the first two terms are of order
\(D^{1/3}/\epsilon^{4/3}\) and the third is \(\sqrt n/\epsilon\). For
\(\epsilon<D^{-1/2}\), we have \(d=\sqrt D/\epsilon\ge D=h\), which gives
\(\widetilde{O}(\sqrt D/\epsilon)\).

\subsection{Related works}

In the classical sampling model, a tester observes independent samples from an unknown
distribution. In an \(\ell_1\) test, one distinguishes membership in a specified family
from \(\ell_1\)-distance at least \(\epsilon\) to every
member~\cite{canonne-survey-2020}. Identity testing takes the reference distribution as
known. For a uniform reference, Paninski~\cite{paninski-uniformity-2008} analyzed a
statistic that counts the outcomes observed exactly once. In two-sample closeness
testing, the distributions \(p\) and \(q\) on \([n]\) are both unknown, and the tester
must distinguish \(p=q\) from \(\|p-q\|_1\ge\epsilon\). At fixed accuracy, Batu et
al.~\cite{batu-et-al-closeness-2013} gave a tester using \(O(n^{2/3}\log n)\) samples.
Chan et al.~\cite{chan-et-al-closeness-2014} determined the sample complexity to be
\(\Theta\big(\max\{n^{2/3}/\epsilon^{4/3},\sqrt{n}/\epsilon^2\}\big)\) samples from
each distribution.

For independence testing on \([n]\times[m]\), the reference distribution
\(q=p_I\times p_J\) depends on the unknown joint distribution \(p\). Batu et
al.~\cite{batu-et-al-independence-2001} related \(\|p-q\|_1\) to the distance from the
nearest product. Acharya, Daskalakis, and Kamath~\cite{acharya-daskalakis-kamath-2015}
used a \(\chi^2\) statistic to compare the joint distribution with a product of
estimated marginals. On \([n]\times[n]\) their tester uses \(O(n/\epsilon^2)\) samples.
Diakonikolas and Kane~\cite{diakonikolas-kane-2016} used marginal flattening to obtain
the optimal rectangular sample bound
\(\Theta\big(\max\{
       n^{2/3}m^{1/3}/\epsilon^{4/3},
       \sqrt{nm}/\epsilon^2\}\big)\).

Bravyi, Harrow, and Hassidim~\cite{bravyi-harrow-hassidim-2011} considered coherent
queries to a function that maps a uniformly random seed to an outcome. At fixed
accuracy, their algorithms test uniformity with \(O(n^{1/3})\) queries and estimate the
\(\ell_1\) distance between two unknown distributions with \(O(\sqrt{n})\) queries.
Montanaro~\cite{montanaro-monte-carlo-2015} improved the dependence on the additive
error by using quantum mean estimation. Under the purified access of
\defn{purified-query-access}, Gily\'en and Li~\cite{gilyen-li-2020} gave an
\(\ell_1\)-closeness tester using \(\widetilde{O}(\sqrt{n}/\epsilon)\) queries. They also reduced independence testing to closeness testing on \(nm\) outcomes, obtaining a query bound of \(\widetilde{O}(\sqrt{nm}/\epsilon)\). For \(\ell_1\)-closeness testing, Luo,
Wang, and Li~\cite{luo-wang-li-2024} also obtained a query bound of \(O(\sqrt{n}/\epsilon)\) and this
gives \(O(\sqrt{nm}/\epsilon)\) a query bound for independence testing by the same reduction.

\subsection{Organization}
\sec{independence-tester} presents the pseudocode in \algo{independence-tester} and
describes marginal flattening, the signs recorded from preliminary samples, and the
checks on new test samples. Subsection~\ref{subsec:tester-decision} gives the decision
rule in terms of the agreement and disagreement counts. \sec{agreement-counts} proves
\thm{independence}. Subsection~\ref{subsec:analysis-correctness} bounds the probability
of rejection under independence and of acceptance under the alternative.
Subsection~\ref{subsec:analysis-queries} derives the total query bound by analyzing the
sampling steps and the quantum comparisons that query the shared marginal lists.

\medskip
\section{The Independence Tester}\label{sec:independence-tester}

At the start of each round, the tester uses ordinary samples from \(p\) to split the
row and column labels. On the refined domain it records the first sign observed at each
pair and applies the marginal checks to new test samples. The resulting agreement and
disagreement counts determine whether the round passes. We retain the notation
\(D=nm\), \(d\), and \(h\) from the introduction and set \(\nu=h/D\).
\algo{independence-tester} specifies the sample sizes, thresholds, and comparison
accuracy.

\begin{algorithm}[!htbp]
\small
\DontPrintSemicolon
\caption{Independence tester}\label{algo:independence-tester}
\Input{Dimensions \(2\le m\le n\), accuracy \(0<\epsilon\le1\), and purified access to \(p\) on \([n]\times[m]\) via \(U_p,U_p^\dagger\) and their controlled versions.}
\Output{\textsc{accept} for \(p=p_I\times p_J\) or \textsc{reject} for \(\|p-p_I\times p_J\|_1\ge\epsilon\), with success probability at least \(2/3\).}
\(D\gets nm,\quad d\gets\min\{\epsilon^{-2},\sqrt D/\epsilon\},\quad
h\gets\min\{D,\lceil(Dd^2)^{1/3}\rceil\},\quad \nu\gets h/D\)\;
\(s_n\gets\min\{h,n\},\quad s_m\gets\min\{h,m\},\quad
K\gets1024,\quad N_{\mathrm{pre}}\gets1024h\)\;
\(N\gets\lceil256h\ln(2^{30}D)\rceil,\quad
b\gets\lceil2^{120}d^2/\nu\rceil,\quad g\gets\lceil d\rceil\)\;
\(M\gets\lceil2^{40}b\nu\rceil,\quad \tau\gets b\nu/(2^{16}d),\quad
\kappa\gets\lceil n/m\rceil,\quad L_*\gets n+\kappa m\)\;
\(A^2\gets\max\{n,D/h\},\quad W_0\gets2^{30}(1+\ln(2D))(M+1)A^2/L_*\)\;
\(q_{\max}\gets1+\lceil\log_2 b\rceil,\quad \delta_{\mathrm{all}}\gets1/100,
\quad \delta\gets\delta_{\mathrm{all}}/(2Kq_{\max}),\quad
r_{\mathrm{cmp}}\gets2\lceil9\ln(1/\delta)\rceil+1\)\;
\BlankLine
\For{\(K\) independent rounds}{
  Draw \(N\) samples from \(p\). Compute row counts \(X_i\) and column counts \(Y_j\)\;
  \(a_i\gets1+\lfloor s_nX_i/N\rfloor\) for \(i\in[n]\),\quad
  \(b_j\gets1+\lfloor s_mY_j/N\rfloor\) for \(j\in[m]\)\;
  \(n'\gets\sum_i a_i,\quad m'\gets\sum_j b_j,\quad u_z\gets1/(n'm')\) for every refined pair \(z\)\;
  Sample \(\widetilde p\) by refining \((i,j)\sim p\) with independent uniform copies
  \(\alpha\in[a_i]\) and \(\beta\in[b_j]\). Sample \(\widetilde q=\widetilde p_I\times\widetilde p_J\)
  using the row and column from two independent refined samples\;
  To generate a signed sample, choose \(k\) uniformly from \(\{1,2,3,4\}\).
  Take \((P,\sigma)\) to be entry \(k\) of the ordered list
  \(\big((\widetilde p,+1),(\widetilde q,-1),(u,+1),(u,-1)\big)\)
  and generate \((Z,\sigma)\) with \(Z\sim P\)\;
  Initialize the table \(G\) with default value \(0\)\;
  \For{\(\ell=1,\ldots,N_{\mathrm{pre}}\)}{
    Generate a signed sample \((z,\sigma)\). If \(G(z)=0\), set \(G(z)\gets\sigma\)\;
  }
  \BlankLine
  Use independent virtual lists \((T_t)_{t=1}^b\) of new signed samples,
  \((R_a)_{a=1}^n\) from \(\widetilde p_I\), and \((C_a)_{a=1}^m\) from \(\widetilde p_J\)\;
  \ForEach{\(t\in[b]\) and axis \((A,L)\in\{(I,n),(J,m)\}\), independently}{
    With probability \(s_L/L\), set \(P_t^A\gets\varnothing\).
    Otherwise choose a uniform permutation \(\pi\) of \([L]\) and an independent
    uniform \(k\in\{0,\ldots,L\}\), and set \(P_t^A\gets\{\pi(1),\ldots,\pi(k)\}\)\;
  }
  For \(T_t=(Z_t,\sigma_t)\), \(Z_t=(r_t,c_t)\), let
  \(B_t=\mathbf 1\{r_t\notin R[P_t^I],\,c_t\notin C[P_t^J]\}\),
  where brackets select entries. Define \(S_t=B_tG(Z_t)\sigma_t\) and
  \(N_s=\#\{t\in[b]:S_t=s\}\) as the target of the comparisons\;
  Use \(2b\) additional automatic covers and \(\kappa\) aliases per column address.
  Let \(W_s\) be the charge in \eqref{eq:count-charge}\;
  A base comparison applies Lemma~\ref{lem:one-sided-count} to \(3b\) words at \(t_0=3b-v-g\).
  It returns \(\mathsf{low}\) for \(N_s\le v,\ W_s\le W_0\), and
  \(\mathsf{high}\) for \(N_s\ge v+g\), with probability at least \(2/3\)
  in each case. Otherwise either answer is allowed\;
  \(\mathsf{Compare}(s,v)\) returns the majority of \(r_{\mathrm{cmp}}\) independent
  base comparisons using the same \(G,T,R,C\) and address choices\;
  \BlankLine
  \ForEach{\(s\) in \((+1,-1)\)}{
    \(L_0\gets0,\quad H_0\gets b\)\;
    \If{\(M+g\le b\)}{
      \lIf{\(\mathsf{Compare}(s,M)=\mathsf{high}\)}{\Return{\textsc{reject}}}
      \(H_0\gets M+g-1\)\;
    }
    \While{\(H_0-L_0>2g\)}{
      \(v\gets\lfloor(L_0+H_0-g)/2\rfloor\)\;
      If \(\mathsf{Compare}(s,v)=\mathsf{low}\), set \(H_0\gets v+g-1\).
      Otherwise set \(L_0\gets v+1\)\;
    }
    \(\widehat N_s\gets(L_0+H_0)/2\)\;
  }
  \lIf{\(\lvert\widehat N_+-\widehat N_-\rvert>\tau\)}{\Return{\textsc{reject}}}
}
\Return{\textsc{accept}}\;
\end{algorithm}

\subsection{Flattening the marginal distributions}\label{subsec:tester-flattening}

To obtain the \(N=O(h\log D)\) ordinary samples specified in
\algo{independence-tester}, measure the two outcome registers of \(U_p\) on each
preparation. Let \(X_i\) count occurrences of row \(i\) and let \(Y_j\) count
occurrences of column \(j\) in the same samples. For an axis of original size \(L\),
set \(s_L=\min\{h,L\}\). For row \(i\) and column \(j\), define the numbers of copies
by
\[
 a_i=1+\left\lfloor\frac{s_nX_i}{N}\right\rfloor,\qquad
 b_j=1+\left\lfloor\frac{s_mY_j}{N}\right\rfloor .
\]
These numbers remain fixed throughout the round. For each subsequent sample \((i,j)\),
choose \(\alpha\in[a_i]\) and \(\beta\in[b_j]\) independently and uniformly to obtain
the refined outcome pair \(z=((i,\alpha),(j,\beta))\).

Write \(\widetilde p\) for the refined joint distribution and
\(\widetilde q=\widetilde p_I\times\widetilde p_J\) for the product of its marginals.
Each copy of \((i,j)\) has probabilities
\[
 \widetilde p_z=\frac{p_{ij}}{a_i b_j},\qquad
 \widetilde q_z=\frac{p_I(i)}{a_i}\,\frac{p_J(j)}{b_j}.
\]
Summing the absolute differences first over the \(a_i b_j\) copies and then over
\((i,j)\) gives \(\|\widetilde p-\widetilde q\|_1=\|p-p_I\times p_J\|_1\). The refined
domain has \(n'=\sum_i a_i\le n+s_n=O(n)\) rows and \(m'=\sum_j b_j\le m+s_m=O(m)\)
columns. Its uniform distribution \(u\) assigns each pair mass \(u_z=1/(n'm')\).

Consider a label of marginal mass \(x>1/s_L\), and let \(X\) count its occurrences in
the \(N\) samples. Hoeffding's binomial inequality~\cite[Eq.~(2.1)]{hoeffding-1963}, applied
to the indicators that a sample misses this label, bounds the lower tail of \(X\). For
the choice of \(N\) in \algo{independence-tester}, \(X/N\) is at least a constant
fraction of \(x\) except with probability \(O(D^{-3})\). On this event the label has at
least \(s_LX/N\) copies, each of mass \(O(1/s_L)\). Labels of mass \(x\le1/s_L\)
already satisfy this bound. Since \(n+m\le D\), a union bound shows that all refined
row masses are \(O(1/s_n)\) and all refined column masses are \(O(1/s_m)\) with
probability \(1-O(D^{-2})\).

\subsection{Generating preliminary samples and signs}\label{subsec:tester-signs}

With the split fixed, generate an outcome pair \(Z\) and a sign \(\sigma\) by choosing
a fresh index \(k\in\{1,2,3,4\}\) uniformly. The four indices correspond to
\((\widetilde p,+1)\), \((\widetilde q,-1)\), \((u,+1)\), and \((u,-1)\), respectively.
To sample from \(\widetilde p\), retain both labels from a single refined sample. To
sample from \(\widetilde q\), take the row from one refined sample and the column from
a second, independent refined sample. To sample from \(u\), choose independently and
uniformly among the \(n'\) refined row labels and the \(m'\) refined column labels.

For each refined outcome pair \(z\), the joint probabilities of the two signs are
\[
 \Pr(Z=z,\sigma=+1)=\frac{\widetilde p_z+u_z}{4},\qquad
 \Pr(Z=z,\sigma=-1)=\frac{\widetilde q_z+u_z}{4}.
\]
It follows that the probability of observing \(z\) and the conditional mean of its sign
are, respectively,
\[
 w_z=\frac{\widetilde p_z+\widetilde q_z+2u_z}{4},\qquad
 \mu_z=\E[\sigma\mid Z=z]
      =\frac{\widetilde p_z-\widetilde q_z}
             {\widetilde p_z+\widetilde q_z+2u_z}.
\]
The uniform samples give each pair a positive probability of appearing. Their
contributions to the numerator of \(\mu_z\) cancel.

Generate \(N_{\mathrm{pre}}=O(h)\) independent preliminary samples. Initialize the
table \(G\) to its default value \(0\). On the first visit to \(z\), store its sign as
\(G(z)\). The table is classical, so a reversible lookup requires no further queries to
\(U_p\). Conditional on the split, the probability that \(z\) receives a sign is
\(1-(1-w_z)^{N_{\mathrm{pre}}}\). Given a visit, the first recorded sign has mean
\(\mu_z\).

\subsection{New test samples and marginal checks}\label{subsec:tester-checks}

The list \(T\) consists of \(b=O(d^2/\nu)\) independent new test samples generated as
in Subsection~\ref{subsec:tester-signs}. Write \(T_t=(Z_t,\sigma_t)\), with
\(Z_t=(r_t,c_t)\) denoting the refined row and column. The round also uses a list
\(R=(R_1,\ldots,R_n)\) of independent samples from \(\widetilde p_I\) and a list
\(C=(C_1,\ldots,C_m)\) of independent samples from \(\widetilde p_J\). Conditional on
the split, the preliminary samples and the three lists \(T,R,C\) are mutually
independent. The lengths of \(R\) and \(C\) are the original axis sizes \(n\) and
\(m\), respectively.

For each new test sample \(t\), choose the addresses on the two axes independently. On
an axis of size \(L\), select the empty set with probability \(s_L/L\). Otherwise
shuffle the \(L\) addresses uniformly and take the first \(k\), choosing \(k\)
independently and uniformly from \(\{0,\ldots,L\}\). Write \(P_t^I\) and \(P_t^J\) for
the resulting sets. These choices require only classical randomness, the axis sizes,
and the skip probabilities. Choose the sets independently across \(t\) and use the same
sets for both counts. The indicator that both checks pass is
\[
 B_t=
 \mathbf 1\{R_a\ne r_t\text{ for every }a\in P_t^I\}\,
 \mathbf 1\{C_a\ne c_t\text{ for every }a\in P_t^J\}.
\]
A check passes automatically when its address set is empty. Each check compares the
full refined label, including its copy index.

To compute the passing probability, fix a marginal list in which the refined label
occurs \(\ell\) times. Taking the first \(k\) entries of the shuffled list is
equivalent to placing a marker uniformly in one of its \(L+1\) gaps and taking the
entries before the marker. The label passes the unskipped check precisely when the
marker precedes all \(\ell\) occurrences. By symmetry this event has probability
\(1/(\ell+1)\). Including the probability of skipping the check gives the passing
probability
\[
 \frac{s_L}{L}+\left(1-\frac{s_L}{L}\right)\frac{1}{\ell+1}.
\]

The three lists of lengths \(b,n,m\) admit queries by address. To implement a read with
\(U_p\), reserve a private register initialized to zero for each address. Let \(V_a\)
prepare the entry at address \(a\), retaining the source choice, copy indices, and
unused registers so that the preparation can be reversed. Apply \(V_a\) to the
addressed register, add its encoded entry to the response register by bitwise XOR, and
apply \(V_a^\dagger\). Keep the private registers throughout the round. As in the
sampling oracle for independent entries of Czajkowski et
al.~\cite[Definition~6(iii)]{czajkowski-et-al-lazy}, the joint preparation factors as
\(\mathcal V=\bigotimes_a V_a\). If \(O\) reads a prepared entry, this sequence
implements \(\mathcal V^\dagger O\mathcal V\) even for a superposition of addresses,
since the factors at unselected addresses cancel. On the algorithm's registers,
successive reads are therefore equivalent to preparing the three lists once and
querying those same lists throughout the round. Each read, including uncomputation,
uses \(O(1)\) purified queries.

\subsection{Counting sign agreements}\label{subsec:tester-counts}

On the same new test samples, define the score and the two counts by
\begin{equation}\label{eq:tester-counts}
 S_t=B_t\,G(Z_t)\,\sigma_t,\qquad
 N_+=\sum_{t=1}^b\mathbf 1\{S_t=+1\},\qquad
 N_-=\sum_{t=1}^b\mathbf 1\{S_t=-1\}.
\end{equation}
For example, a new test sample of sign \(-1\) agrees with a recorded sign of \(-1\) and
contributes one agreement if both marginal checks pass. If either check fails or the
pair has no recorded sign, then \(S_t=0\).

Fix \(G,R,C\) and the split. Let \(\alpha_R(r)\) and \(\alpha_C(c)\) denote the passing
probabilities from Subsection~\ref{subsec:tester-checks} for row \(r\) and column
\(c\). Taking the expectation over a new test sample and its address choices gives
\[
 \E[S_t\mid G,R,C,\text{split}]
   =\frac14\sum_{z=(r,c)}
      G(z)\,\alpha_R(r)\alpha_C(c)\,(\widetilde p_z-\widetilde q_z).
\]
Under independence the sign is fair conditional on the outcome pair. Consequently
\(N_+-N_-\) has mean zero even after fixing all address choices.

For sign \(s\), a new test sample is eligible if its pair has a recorded sign and
\(G(Z_t)\sigma_t=s\). An ineligible sample is covered automatically. An eligible sample
is covered if an inspected row or column entry matches its corresponding label. Thus
\(N_s\) counts the uncovered samples. We append \(2b\) public, automatically covered
indices for the counting procedure.

The reference array contains one entry per row address and \(\kappa=\lceil n/m\rceil\)
aliases per column address, for a total length of \(L_*=n+\kappa m\). All aliases of a
column address refer to the same word. For each eligible sample \(t\), let \(H_t^*\)
denote the number of certifying addresses. Each selected row match contributes one and
each selected column match contributes \(\kappa\). Define the charge and its public cap
by
\begin{equation}\label{eq:count-charge}
 W_s=\sum_{\substack{t\ {\rm eligible\ for}\ s\\H_t^*>0}}\frac1{H_t^*},
 \qquad
 W_0=\frac{2^{30}(1+\ln(2D))(M+1)A^2}{L_*},\qquad
 A^2=\max\{n,D/h\}.
\end{equation}
Automatic covers have zero charge. The tester uses the public cap \(W_0\), and the
analysis bounds the realized charge \(W_s\) over the sampling randomness.

For a threshold \(v\), the base comparison has the following guarantees.
\[
 \begin{array}{ll}
 N_s\le v\ \text{and}\ W_s\le W_0 &\Longrightarrow\ \mathsf{low},\\
 N_s\ge v+g &\Longrightarrow\ \mathsf{high}.
 \end{array}
\]
Each promised answer is correct with probability at least \(2/3\). Either answer is
allowed on the remaining inputs. Lemma~\ref{lem:one-sided-count} implements the
comparison by querying the new test samples and the shared references.

Suppose \(W_s\le W_0\). A valid low answer implies \(N_s\le v+g-1\), and a valid high
answer implies \(N_s\ge v+1\). The updated interval therefore contains \(N_s\) even
when the count lies inside the threshold gap. As soon as \(H_0-L_0\le2g\), the midpoint
satisfies \(\lvert\widehat N_s-N_s\rvert\le g\).

When an update changes the interval width from \(w\) to \(w'\), either answer gives
\(w'-g\le(w-g)/2\). Thus at most \(q_{\max}\) calls are made per sign, including the
overflow test, even on data outside the promise. Fix the data and condition on all
preceding answers. By Hoeffding's inequality~\cite[Theorem~1]{hoeffding-1963} applied
to the error indicators of the \(r_{\mathrm{cmp}}\) independent base comparisons, the
majority gives an incorrect required answer with probability at most
\(\exp(-r_{\mathrm{cmp}}/18)\le\delta\). A union bound over at most \(2Kq_{\max}\)
adaptive calls bounds the probability of any incorrect \emph{required} answer by
\(\delta_{\mathrm{all}}\).

\subsection{Decision rule}\label{subsec:tester-decision}

In each round, \algo{independence-tester} checks the agreement count \(N_+\) and then
the disagreement count \(N_-\). For each sign \(s\), if \(M+g\le b\), the tester first
compares \(N_s\) with the overflow threshold \(M=O(b\nu)\). A valid comparison returns
low if \(N_s\le M\) and \(W_s\le W_0\), high if \(N_s\ge M+g\), and either answer
otherwise. On a high answer the tester rejects immediately. On a low answer it begins
the count search on \([0,M+g-1]\). If \(M+g>b\), the search starts on \([0,b]\) without
an overflow check.

If both searches return estimates, let \(\widehat\Delta=\widehat N_+-\widehat N_-\).
The tester rejects if \(\lvert\widehat\Delta\rvert>\tau\), where \(\tau=O(b\nu/d)\). If
\(\lvert\widehat\Delta\rvert\le\tau\), the round passes. The tester accepts after all
\(K\) independent rounds pass.

Suppose \(W_+,W_-\le W_0\). If the comparisons give valid answers and both searches
return estimates, then
\[
 \lvert\widehat\Delta-\Delta\rvert
 \le\lvert\widehat N_+-N_+\rvert+\lvert\widehat N_--N_-\rvert\le2g,
 \qquad \Delta=N_+-N_- .
\]
Let \(E_{\mathrm{pass}}\) denote the event that both charges are at most \(W_0\), that
\(\lvert\Delta\rvert\le\tau-2g\), and that both counts are at most \(M\) whenever the
overflow test is used. Let \(E_{\mathrm{reject}}\) be the event that at least one of
the following conditions holds.
\[
 \max\{N_+,N_-\}\ge M+g,\qquad\text{or}\qquad
 \max\{W_+,W_-\}\le W_0\ \text{ and }\ \lvert\Delta\rvert>\tau+2g .
\]
The first condition forces rejection at the overflow test for any values of the
charges. If the tester examines the other sign first, that search either rejects or
finishes within its prescribed number of calls. The second condition forces rejection
either at overflow or by the signed difference. Thus, whenever the required answers are
correct, the round passes on \(E_{\mathrm{pass}}\) and the tester rejects on
\(E_{\mathrm{reject}}\).

Under independence, valid comparisons can lead to rejection only if at least one round
lies outside \(E_{\mathrm{pass}}\). Under the alternative, valid comparisons can lead
to acceptance only if every round misses \(E_{\mathrm{reject}}\). A union bound in the
first case and independence of the round data in the second give
\begin{align*}
 \Pr_p(\text{\textsc{reject}})
   &\le \delta_{\mathrm{all}}+K\,\Pr_p(E_{\mathrm{pass}}^{\,c})
   &&\text{if } p=p_I\times p_J,\\
 \Pr_p(\text{\textsc{accept}})
   &\le \delta_{\mathrm{all}}+
          \bigl(1-\Pr_p(E_{\mathrm{reject}})\bigr)^K
   &&\text{if } \|p-p_I\times p_J\|_1\ge\epsilon .
\end{align*}
It remains to bound each right-hand side by \(1/3\) to obtain the success probability
in \thm{independence}.

\medskip
\section{Analysis of the Independence Tester}\label{sec:agreement-counts}

\subsection{Correctness}\label{subsec:analysis-correctness}

Write \(H=N_{\mathrm{pre}}\), \(\rho_z=\widetilde p_z-\widetilde q_z\), and
\(F_z=\alpha_R(r)\alpha_C(c)\) for \(z=(r,c)\). Let \(\mathcal F\) denote the data
consisting of the split, the preliminary table \(G\), and the marginal lists \(R,C\).
Conditional on \(\mathcal F\), each new test sample and its address choices are
independent of those for every other index \(t\). Define
\[
 \pi_s=\Pr(S_t=s\mid\mathcal F),\qquad
 Y=\pi_+-\pi_-=\tfrac14\sum_z G(z)F_z\rho_z .
\]
It follows that \(N_s\) is conditionally binomial with parameters \(b,\pi_s\), and that
\(\Delta=N_+-N_-\) has conditional mean \(bY\) and variance at most \(b(\pi_++\pi_-)\).

\emph{The probability of satisfying the counting promise.}
Fix \(\mathcal F\). Suppose a label occurs \(j\) times in a marginal list of length
\(L\). In an unskipped prefix the number of matches is uniform on \(\{0,\ldots,j\}\),
since the cutoff is equally likely to occupy any position relative to the \(j\)
occurrences. Let \(J_L\) count the inspected matches, with \(J_L=0\) when the check is
skipped. Interpreting \(1/J_L\) as zero when \(J_L=0\), we bound its expectation by
\((1+\ln(2D))/(1+j)\) using the harmonic sum. For an eligible sample at \(z\), the
number of certifying addresses is \(J_n+\kappa J_m\). Under the same zero convention
its reciprocal is at most \(1/J_n+1/(\kappa J_m)\). Put \(a_L=s_L/L\). If the row and
column occur \(j_r,j_c\) times, respectively, then \(F_z\ge a_m/(1+j_r)\) and
\(F_z\ge a_n/(1+j_c)\). Multiplying these bounds by the probability that a new test
sample occurs at the given pair and is eligible, then summing over pairs and the \(b\)
samples, gives
\begin{align}
 L_*\E[W_s\mid\mathcal F]
 &\le 3(1+\ln(2D))b\pi_s(n/a_m+m/a_n) \notag\\
 &\le 6(1+\ln(2D))b\pi_s A^2 .\label{eq:mean-witness-charge}
\end{align}
Here we used \(L_*\le3n\), \(L_*/\kappa\le3m\), and \(A^2=\max\{n,D/h\}\). In
particular, whenever \(b\pi_s\le2M\), Markov's inequality gives
\begin{equation}\label{eq:charge-exception}
 \Pr(W_s>W_0\mid\mathcal F)\le 12/2^{30}.
\end{equation}
This event depends only on the round data. Its probability is bounded once because
every threshold for that sign uses the same \(W_s\).

\emph{Independent inputs.}
Call a split good if every refined label on an axis of original size \(L\) has marginal
mass at most \(2/s_L\). For an original label of mass \(x>1/s_L\), the empirical count
is at least \(Nx/2\) except with probability \(e^{-Nx/8}\). On this event each copy has
mass at most \(2/s_L\). The remaining labels already satisfy the bound. Taking a union
bound over labels gives a probability of at most \(D(2^{30}D)^{-32}<2^{-30}\) for a bad
split.

Under independence, \(Y=0\) for every \(\mathcal F\). Fix a good split and the table
\(G\), and take expectations over the marginal lists. If \(J\sim\mathrm{Bin}(L,x)\)
counts the occurrences of a refined label of mass \(x\) in its list, then
\(x\,\E[1/(1+J)]=(1-(1-x)^{L+1})/(L+1)\). Multiplying the label mass by its expected
passing probability gives at most \(xa_L+1/(L+1)\le3/L\). The two marginal lists are
independent, so \(\widetilde q_z\,\E F_z\le9/D\). We also have \(u_zF_z\le1/D\). Under
the null, \(w_z=(\widetilde q_z+u_z)/2\), and \(G\) contains at most \(H\) recorded
pairs. Hence
\[
 \E(\pi_++\pi_-)\le 5H/D,\qquad
 \E(N_++N_-)\le5bH/D,\qquad
 \E\Delta^2\le5bH/D .
\]
Apply Markov's inequality to the last two expectations, and apply
\eqref{eq:mean-witness-charge} to bound the mean of \(W_++W_-\). Substituting
\(H=1024h\), \(b\nu\ge2^{120}d^2\), and \(\tau-2g\ge\tau/2\) and including the
probability of a bad split gives
\[
 \Pr_p(E_{\mathrm{pass}}^c)\le
 2^{-30}+\frac{5bH}{DM}
 +\frac{5bH}{D(\tau-2g)^2}
 +\frac{30bH}{2^{30}D(M+1)}<2^{-20}.
\]
The probability of rejection is therefore at most
\(\delta_{\mathrm{all}}+K2^{-20}<1/3\).

\emph{Dependent inputs.}
Write \(\ell=\sum_z|\rho_z|\ge\epsilon\). We first show that
\begin{equation}\label{eq:promised-signal}
 \Pr_p\{|bY|\ge4\tau\}\ge1/8.
\end{equation}
Fix a good split and suppose \(\epsilon\ge D^{-1/2}\). Let \(K_z\) count the
preliminary occurrences of \(z\), write \(E=\{z:K_z>0\}\) for the set of observed
pairs, and put
\[
 X=\sum_z K_z|\rho_z|/w_z,\qquad
 J_*=\sum_{z\in E}K_z^2/(w_z F_z).
\]
The \(H\) preliminary draws are independent and \(|\rho_z|/w_z\le4\), so \(\E X=H\ell\)
and \(\operatorname{Var}(X)\le4H\ell\). Since \(h\ge d=\epsilon^{-2}\), we have
\(H\ell\ge1024\). It follows from Chebyshev's inequality that, with probability at
least \(1-1/64\), \(X\ge H\ell/2\).

For refined marginal masses \(r_i,c_j\), put \(B_{ij}=(1+nr_i)(1+mc_j)\). On each axis,
the reciprocal of the passing probability is at most one plus the number of occurrences
of the label in the corresponding list. It follows that \(\E[1/F_{(i,j)}]\le B_{ij}\),
\(\sum_{ij}B_{ij}\le9D\), and \(\max_{ij}B_{ij}\le9D/h\). The last bound uses the good
split and \(s_ns_m\ge h\). The count \(K_z\) is binomial and independent of the
marginal lists. Its second moment therefore gives \(\E J_*\le9DH+9DH^2/h\le18DH^2/h\).
With probability at least \(47/64\), we have both \(X\ge H\ell/2\) and
\(J_*\le72DH^2/h\). On this event the Cauchy--Schwarz inequality gives
\begin{equation}\label{eq:coarse-bias}
 \sum_{z\in E}F_z\rho_z^2/w_z\ge X^2/J_*\ge \nu\ell^2/288 .
\end{equation}

Condition on the split, the preliminary outcome pairs, and both marginal lists, and
write \(\E_*\) for the resulting expectation. The first observed signs are independent,
with respective means \(\rho_z/(4w_z)\). Each summand of \(Y\) has nonnegative mean and
fixed squared magnitude. Consequently
\[
 \E_*Y=\tfrac1{16}\sum_{z\in E}F_z\rho_z^2/w_z,\qquad
 \E_*Y^2\ge\max\{(\E_*Y)^2,\tfrac1{16}\sum_{z\in E}F_z^2\rho_z^2\}.
\]
We also have \(\E_*Y^4\le3(\E_*Y^2)^2\). To verify this bound, expand the fourth power
and consider the four summands in each product. If their indices are distinct, each of
the three pairings gives a product of pair expectations equal to the expectation of the
original product. If an index is repeated, pair two equal indices. Their product is
deterministic, so this pairing again gives equality. All other pairings give
nonnegative contributions. Summing over indices proves the fourth-moment bound. The
Paley--Zygmund inequality~\cite{paley-zygmund-1932} applied to \(Y^2\) at one quarter
of its conditional mean now gives \(\lvert Y\rvert\ge\frac12\sqrt{\E_*Y^2}\) with
probability at least \(3/16\) whenever \(\E_*Y^2>0\). On the event in
\eqref{eq:coarse-bias}, the threshold in this probability estimate is at least \(\nu/(9216d)>4\tau/b\). Multiplying the conditional probability by
the probability of a good split and by \(47/64\) proves \eqref{eq:promised-signal}.

In the remaining regime, \(\epsilon<D^{-1/2}\), we have \(h=D\), \(F_z=1\), and
\(w_z\ge1/(8D)\). The preliminary samples miss any given pair with probability at most
\(e^{-128}\). The expected squared discrepancy on unvisited pairs is thus at most
\(e^{-128}\|\rho\|_2^2\). By Markov's inequality, with probability at least
\(1-2e^{-128}\) the visited pairs account for at least half of \(\|\rho\|_2^2\). On
this event
\[
 \tfrac12\sqrt{\E_*Y^2}\ge
 \|\rho\|_2/(8\sqrt2)\ge \epsilon/(16\sqrt{2D})>4\tau/b .
\]
Applying the same conditional bound for the recorded signs proves
\eqref{eq:promised-signal}.

Finally, fix \(\mathcal F\) satisfying \(|bY|\ge4\tau\). If \(b\pi_s>2M\) for either
sign, the binomial lower-tail bound and \(g\le M/4\) imply \(N_s\ge M+g\) except with
probability at most \(e^{-M/16}\). The large-count guarantee holds for every value of
\(W_s\). Otherwise, \eqref{eq:charge-exception} applies to both signs and
\(\operatorname{Var}(\Delta\mid\mathcal F)\le4M\). Chebyshev's inequality gives
\(\Pr(|\Delta-bY|\ge\tau\mid\mathcal F)\le4M/\tau^2\). Since \(M\le2^{41}b\nu\), the
sum of this probability and the probabilities of the two charge exceptions is less than
\(1/16\). Thus \(\Pr_p(E_{\mathrm{reject}})\ge(1/8)(15/16)>1/16\). By the second error
bound in Subsection~\ref{subsec:tester-decision}, the probability of acceptance is at
most \(\delta_{\mathrm{all}}+(15/16)^K<1/3\).

\subsection{Query Complexity}\label{subsec:analysis-queries}

The comparison used by the tester queries two arrays, one of raw words and one of
reference words. A raw word is covered automatically if its index and value satisfy a
public predicate. Otherwise a reference address certifies the raw word if the address
and its value satisfy a second public predicate for that word and index. The word is
covered if it has at least one such certifying address. In our application, the
predicates use the fixed preliminary table and the selected sets of addresses.

\begin{lemma}\label{lem:one-sided-count}
Let \(B\) be the number of raw words and \(L\) the length of the reference array. Fix
\(B/2\le t<t+g\le B\) and \(W_0>0\). Consider the problem of distinguishing at most
\(t\) covers from at least \(t+g\) covers. In the latter case assume
\(\sum_i1/H_i\le W_0\), where the sum runs over covered words that are not automatic
and \(H_i\) counts the addresses certifying word \(i\). With \(\xi=1+(B-t)/g\), this
comparison can be made with bounded error using
\[
 \widetilde{O}\!\left(\sqrt{B\xi/g}+\sqrt{LW_0/g}\right)
\]
word queries.
\end{lemma}

\begin{proof}
We construct a dual adversary witness whose vectors are indexed by an input and a
queried address. For any two inputs with opposite outputs, the inner products must sum
to one over the addresses where the inputs differ. By the general adversary
characterization of Lee et al.~\cite[Theorem~1.1]{lee-et-al-2011}, the maximum over
inputs of the sum of squared vector norms then bounds the quantum query complexity up
to a constant factor. On a low input, extend the set of covered indices to a set \(A\)
of size \(t\). On a high input, choose \(t+g\) covered indices to form a set \(B'\).
Both choices enter only in the construction of the witness.

We use the gapped group-testing weights of Ambainis et
al.~\cite[Eqs.~(12)--(16)]{ambainis-et-al-ggt}. A group \(S\) of size \(k\) receives
low weight \(\alpha_k\) when \(S\cap A=\varnothing\) and high weight \(\beta_k\) when
\(S\cap B'=\{i\}\). The remaining weights are zero. For a pair of inputs with opposite
outputs, the products of their weights sum to one over the common groups. Write
\(P=B-t\). For \(1\le k\le P-g+1\), the sum of squared weights over groups of size
\(k\) on either input is
\[
 w_k=\sqrt{\frac{t+g}{Pk}\,
        \frac{\binom{P-g}{k-1}}{\binom{P-1}{k-1}}}.
\]
To verify the normalization, note that \(\alpha_k\beta_k=[P\binom{P-1}{k-1}]^{-1}\). If
\(\lambda=|B'\setminus A|\), the sum of products over common groups is
\[
 \sum_{k=1}^{P-\lambda+1}
 \frac{\lambda\binom{P-\lambda}{k-1}}{P\binom{P-1}{k-1}}=1.
\]
Indeed, the \(k\)th summand is the probability that a uniformly shuffled list of the
\(P\) indices outside \(A\) first meets \(B'\) at position \(k\). On a high input,
symmetry makes the distinguished index \(i\) uniform in \(B'\) under the squared
weights. For \(0\le j\le1\) we have
\begin{equation}\label{eq:group-size-moment}
 \sum_k w_k k^j=O\!\left(\sqrt{B/g}\,\xi^j\right).
\end{equation}
To see this, observe that for \(P>1,\ g\ge2\) the binomial ratio is at most
\(\exp(-(k-1)(g-1)/(P-1))\). Summing \(k^{j-1/2}\) against this exponential factor
gives a cutoff at scale \(P/g\). For \(g=1\) we sum directly up to \(P\), and for
\(P=1\) only \(k=1\) contributes.

Choose an integer \(a\ge1\). All groups begin with the same \(a-1\) reference reads,
chosen uniformly with replacement. A vector coordinate specifies the stage, the ordered
addresses and values already read, and the address of the next read. Assign it the
square root of the path probability through the choice of that next address. The
squared norm at each stage is one. For a fixed pair of low and high inputs, let \(e\)
be the probability that the prefix values agree. Grouping the paths by their first
disagreement gives a cross contribution of \(1-e\).

For the remaining stages, each group has separate coordinates for automatic and
certified covers. The low vector uses both sectors. The high vector uses the sector
corresponding to its distinguished index \(i\). In each sector let \(p_0\) denote the
low base probability of an edge and \(p_1\) the high path probability. Following
Belovs's learning-graph normalization~\cite{belovs-learning-graphs-2012}, assign
amplitudes \(\sqrt{p_0}\) and \(p_1/\sqrt{p_0}\) and multiply them by the low and high
group weights, respectively. On a high path we evaluate the base probability using the
same input-independent law for sampling addresses, with the values observed on that
path. At a matching history the cross product is therefore the high flow multiplied by
the two group weights.

In the automatic sector, the low side queries a uniformly chosen raw index in \(S\) and
the high side queries \(i\). On a common group the raw words at \(i\) differ because
\(i\) is uncovered on the low input. Before multiplication by the group weights, the
cross contribution is \(e\). In the certified sector we append a reference read. The
low side chooses it uniformly, and the high side chooses it uniformly from the \(H_i\)
certifiers. We then make the same raw-index query as in the automatic sector. At this
last step we retain the multiset of all \(a\) reference addresses and their values,
including multiplicities, and sum the path probabilities over orders \emph{before}
assigning amplitudes. Let \(\beta\) denote the fraction of \(i\)'s certifiers at which
the reference words agree. The appended edge contributes \(e(1-\beta)\). If the labeled
multisets agree, the reference values agree at every address read. The raw words must
then differ, since equal raw words would also certify \(i\) on the low input. The
terminal edge thus contributes \(e\beta\). Each common group contributes \(e\) times
the product of its two weights. Adding the shared prefix gives \(1-e+e=1\).

We next bound the squared norms. Before multiplication by the outer weights, the low
base vector has squared norm one at each group stage. On the high side the squared norm
is \(k\) at an automatic raw query and \(L/H_i\) at the appended reference query. At
the terminal raw query, let \(J_i\) count the certifying entries in the multiset, with
multiplicity. The density of this multiset on the high side relative to \(a\) uniform
reads is \(J_iL/(aH_i)\), since any of the \(a\) positions could have supplied the
designated certifier. Under uniform reads, \(J_i\sim\mathrm{Bin}(a,H_i/L)\). Its second
moment gives the squared norm on the high side as
\[
 k\,\E\!\left[\left(\frac{J_iL}{aH_i}\right)^2\right]
 =k\left(1+\frac{L/H_i-1}{a}\right).
\]
For fixed \(k\), average over the uniform choice of \(i\in B'\), counting automatic
indices as zero. The resulting certified-sector average of \(L/H_i\) is at most
\(LW_0/(t+g)\le2LW_0/B\). For each size, stage, and sector, scale the low and high
vectors by reciprocal factors. Given squared norm bounds \(U,V\), this yields the bound
\(\sqrt{UV}\) on each side and preserves all cross products. Applying
\eqref{eq:group-size-moment} with \(j=0,1/2\) gives the total bound
\[
 O\!\left(a+\sqrt{B/g}\left[
 \sqrt{\xi}+\sqrt{LW_0/B}+\sqrt{\xi LW_0/(aB)}\right]\right).
\]
Choose \(a=\lceil\xi\rceil\). The inequality \(\xi\le2B/g\) now proves the lemma in the
query model with unrestricted gates and workspace.
\end{proof}

At threshold \(v\), apply Lemma~\ref{lem:one-sided-count} with \(B=3b,\ L=L_*\) and
\(t=3b-v-g\). The \(2b\) padded indices are covered automatically, so there are exactly
\(3b-N_s\) covers. Thus \(N_s\ge v+g\) corresponds to at most \(t\) covers, the case in
which the lemma imposes no charge condition. Every threshold satisfies \(v+g\le b\),
hence \(t\ge2b\). We also have \(v+g\le M+g\) whether or not the initial overflow test
is used. Consequently \(\xi=1+(v+g)/g=O(d)\) by \(b\nu=O(d^2)\) and
\(g=\lceil d\rceil\). The public cap satisfies \(L_*W_0=O((1+\ln(2D))b\nu A^2)\).
Substitution in the lemma bounds the cost of a base comparison by
\[
 \widetilde{O}(\sqrt b+\sqrt d\, A)
   =\widetilde{O}\!\left(d\sqrt{D/h}+\sqrt{nd}\right).
\]
Each alias of a column word addresses the original register. By the simulation in
Subsection~\ref{subsec:tester-checks}, a word read costs \(O(1)\) purified queries,
even for repeated reads or reads in superposition. The tester fixes the cap before
making these queries. Its sampling guarantee enters the error analysis, and the query
bound holds on every execution.

Flattening the marginals and generating the preliminary samples cost
\(\widetilde{O}(h)\) queries. Accounting for the \(K\) rounds, at most \(q_{\max}\)
comparisons per sign per round, and \(r_{\mathrm{cmp}}\) repetitions per comparison
introduces only a polylogarithmic factor. When \(\epsilon\ge D^{-1/2}\), we have
\(d=\epsilon^{-2}\). Both \(h\) and \(d\sqrt{D/h}\) are then of order
\(D^{1/3}\epsilon^{-4/3}\), and the remaining term is \(\sqrt n/\epsilon\). When
\(\epsilon<D^{-1/2}\), we have \(h=D\) and \(d=\sqrt D/\epsilon\ge D\). Each of \(h\),
\(d\sqrt{D/h}\), and \(\sqrt{nd}\) is therefore at most \(d\). These estimates give the
bound in \thm{independence} for the full parameter range.


\ifCLASSOPTIONcaptionsoff
  \newpage
\fi

\end{document}